\documentclass[a4paper,11pt]{article}

\usepackage{acro}
\DeclareAcronym{cr}{
  short=CR,
  long=cosmic ray,
}

\DeclareAcronym{eas}{
  short=EAS,
  long=extensive air shower,
}

\DeclareAcronym{pde}{
  short=PDE,
  long=partial differential equation,
}

\DeclareAcronym{mceq}{
  short=MCEq,
  long=matrix cascade equations,
}

\DeclareAcronym{pmt}{
  short=PMT,
  long=photomultiplier tube,
}

\DeclareAcronym{dom}{
  short=DOM,
  long=digital optical module,
}

\DeclareAcronym{mpe}{
  short=MPE,
  long=multi-photo-electron,
}

\DeclareAcronym{mc}{
  short=MC,
  long=Monte Carlo simulation,
}

\DeclareAcronym{mjd}{
  short=MJD,
  long=modified Julian date,
}

\DeclareAcronym{airs}{
  short=AIRS,
  long=atmospheric infrared sounder,
}

\DeclareAcronym{nasa}{
  short=NASA,
  long=the National Aeronautics and Space Administration,
}

\DeclareAcronym{sso}{
  short=SSO,
  long=sun-synchronous orbit,
}

\DeclareAcronym{pdf}{
  short=PDF,
  long=probability density function,
}

\DeclareAcronym{cdf}{
  short=CDF,
  long=cumulative distribution function,
}

\DeclareAcronym{llh}{
  short=LLH,
  long=log-likelihood,
}

\DeclareAcronym{dof}{
  short=DoF,
  long=degrees of freedom,
}

\DeclareAcronym{fft}{
  short=FFT,
  long=fast Fourier transform,
}

\DeclareAcronym{ecmwf}{
  short=ECMWF,
  long=european centre for medium-range weather forecasts,
}

\usepackage{siunitx}

\usepackage{pos}
\usepackage{lipsum} 

\title{12-Years Observation of Seasonal Variation of Atmospheric Neutrino Flux with IceCube}

\ShortTitle{12-Years Observation of Seasonal Variation of Atmospheric Neutrino Flux with IceCube}

\author{The IceCube Collaboration \\{\normalsize \normalfont(a complete list of authors can be found at the end of the proceedings)}\\}

\emailAdd{shuyang.deng@rwth-aachen.de}

\abstract{

High-energy atmospheric muon neutrinos are detected by the IceCube Neutrino Observatory with a high rate of almost a hundred thousand events per year. Being mainly produced in meson decays in cosmic-ray-induced air showers in the upper atmosphere, the flux of these neutrinos is expected to depend on atmospheric conditions and thus features a seasonal variation. The correlation between temperature fluctuations and variations of the neutrino rates can be described with a slope parameter $\alpha$, whose previous measurement with 6 years of IceCube data indicated a discrepancy to theoretical expectations. In this work, we present an update of the previous analysis, extending the statistics to 12-years of IceCube data, as well as adding a region in the Northern Hemisphere to the analysis. We estimate the slope parameter in the Southern Hemisphere to be $\alpha=0.325\pm0.022$, which confirms the previous observation of the tension between the theoretical predictions and experimental measurements with significance $>3\sigma$. Furthermore, the seasonal variation in the Northern Hemisphere has also been observed for the first time, with $\alpha=0.731\pm0.222$. Investigations into systematic effects reveal that the observations not only show a weaker correlation compared to the predictions, but also deviate from the expected linear relation between the atmospheric neutrino flux and the atmospheric temperature.

\vspace{4mm}

{\bfseries Corresponding authors:}
Shuyang Deng$^{1*}$
\\
{$^{1}$ \itshape RWTH Aachen University}\\
$^*$ Presenter
}

\ConferenceLogo{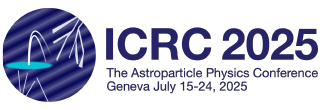}

\FullConference{39th International Cosmic Ray Conference (ICRC2025)\\
 15–24 July 2025\\
Geneva, Switzerland\\}

\begin{document}

\maketitle

\section{Introduction}
Atmospheric neutrinos are generated in \acp{eas}, which are subsequences of \ac{cr} particles interacting with air molecules in the atmosphere of the Earth. They provide a unique window for studying the characteristics of neutrinos themselves, as well as the particle-physics processes they participate in. Furthermore, for large-scale neutrino telescopes, atmospheric neutrinos are usually the dominant irreducible background for astrophysical neutrinos \cite{IceCube_Atm.2011}. For these studies, understanding the atmospheric neutrino flux is an essential requirement.

 The high-energy atmospheric neutrinos detected by IceCube are dominantly produced in the early development of \acp{eas}, as decay products of $K^\pm$ and $\pi^\pm$. However, those charged mesons can also inelastically interact with atmospheric nuclei before they decay, preventing the production of high-energy neutrinos. It is expected that the production of atmospheric neutrinos features a linear dependency on the atmospheric density, derived from the competition between decay and re-interaction \cite{Gaisser_book_2016}.

A well-known analytical approximation for such atmospheric lepton flux is the Gaisser approximation. When applying it to the neutrino flux from pion and kaon decay, the neutrino production yield $P_\nu$, which is the gain of neutrino flux $\Phi_\nu$ at given atmospheric slant depth $X$, neutrino energy $E_\nu$ and local zenith angle $\theta^*$ can be written as \cite{Gaisser_book_2016}
\begin{equation}
    \label{eq:Gaisser}
    \begin{alignedat}{2}
        P_\nu(E_\nu,X,\theta^*)& \equiv \frac{d\Phi_\nu(E_\nu,X,\theta^*)}{dX}
        \\
        &=\Phi_N(E_\nu)\left(\frac{\mathcal{A}_{\pi\nu}(X)}{1+\mathcal{B}_{\pi\nu}(X)\frac{E_\nu\cos(\theta^*)}{\epsilon_\pi(T(X))}}+\frac{\mathcal{A}_{K\nu}(X)}{1+\mathcal{B}_{K\nu}(X)\frac{E_\nu\cos(\theta^*)}{\epsilon_K(T(X))}}\right) \,.
    \end{alignedat}
\end{equation}
Here, $\Phi_N$ is the primary \ac{cr} spectrum. The factors $\mathcal{A}_{i\nu}$ and $\mathcal{B}_{i\nu}$ contain the branching ratio, cross-sections, and kinematic factors from the respective parent mesons, $\epsilon_i$ is the critical energy marking the transfer from the decay-dominating to re-interact-dominating regime of the respective parent meson, and $T(X)$ is the atmospheric temperature profile.
The asterisk on the local zenith angle $\theta$ indicates the correction necessary to account for the curvature of the Earth for $\theta>60^\circ$ \cite{Gaisser:2014eaa}.

The IceCube Neutrino Observatory measures atmospheric neutrinos with energy above 100~\si{GeV} at a rate of hundreds of neutrinos a day \cite{Icecube_munudata_2022}. An experimental measurement of the correlation between atmospheric neutrino flux and the temperature can be made by combining the high-statistic neutrino sample produced by IceCube and a global atmospheric profile dataset. Such an atmospheric measurement is performed by \ac{airs} \cite{aqua_2003}, which is a detector mounted on the Aqua satellite launched by NASA in May 2002 \cite{aqua_2003}. As a spaceborne instrument, it is capable of measuring global atmospheric profiles on a daily basis.

Such a measurement has been successfully done with 6 years of IceCube data \cite{6year_paper}. However, indication towards unresolved systematic effects was also observed. In this work, we will present an updated analysis with 12 years of data and also investigate a new region in the Northern Hemisphere.

\section{Analysis Methods and Previous Results}
IceCube \cite{Aartsen:2016nxy} is a neutrino telescope with a detector volume of about one cubic kilometer, which is achieved by implementing light sensors deep into the Antarctic ice shell.
Since the detector is located in the South Pole, the zenith angle in IceCube is defined so that $0^\circ$ is the vertical down-going direction, corresponding to events originating from right above the detector, and $180^\circ$ is the vertical up-going direction, corresponding to events originating from the geographical North Pole. 
IceCube has successfully measured the seasonal variation of the atmospheric neutrino flux with 6 years of experimental data \cite{6year_paper}. 
The analysis method of this work is based on the framework described in  \cite{6year_paper} and has been further developed to incorporate more data and increase the accuracy. To quantify the correlation between atmospheric temperature and the neutrino rates, we define the slope parameter $\alpha$ as \cite{Gaisser_book_2016}
\begin{equation}
    \label{eq:alpha_local}
    \alpha(E,\theta,\phi) = \frac{T}{\Phi_\nu(E_\nu,\theta,\phi)}\frac{\partial\Phi_\nu(E_\nu,\theta,\phi)}{\partial T} \,.
\end{equation}
where $\theta$ is the zenith angle observed with IceCube and $\phi$ is the azimuth angle. Since a direct measurement of the production height of individual neutrinos is not feasible, we need to integrate the neutrino yield along the slant depth and define the effective temperature $T_\mathrm{eff}$ as

\begin{equation}
    \label{eq:teff}
    T_\mathrm{eff}(\theta,\phi,t) = \frac{\iint A_\mathrm{eff}(E_\nu,\theta)P(E_\nu,X,T,\theta^*(\theta)T(X,\theta,\phi,t)dE_\nu dX}{\iint A_\mathrm{eff}(E_\nu,\theta)P(E_\nu,X,T,\theta^*(\theta))dE_\nu dX}
\end{equation}
using the neutrino yield described in Eq.~\ref{eq:Gaisser}. Here, $A_\mathrm{eff}$ is the effective area, which depends on the data selection. For this work, we use the northern track sample of IceCube. It contains up-going muon tracks, which can only be induced by neutrinos, since muons cannot penetrate the whole earth and reach IceCube from the bottom. The selection and reconstruction process of this sample are described in \cite{Icecube_munudata_2022}. 

We further integrate the effective temperature over chosen regions in which the atmosphere features a significant temperature variation. Thus, for the seasonal variation of the total flux in a given observation region, we expect a linear correlation
\begin{equation}
    \label{eq:alpha}
    \delta R(t)=\alpha \cdot \delta T_\mathrm{eff}(t) \,,
\end{equation}
with the dimensionless relative neutrino rate and effective temperature 
\begin{align}
    \label{eq:delta}
    \delta R(t)      & = \frac{R_\nu(t)-\left<R_\nu\right>}{\left<R_\nu\right>}\\
    \delta T_\mathrm{eff}(t)& = \frac{T_\mathrm{eff}(t)-\left<T_\mathrm{eff}\right>}{\left<T_\mathrm{eff}\right>} \,,
\end{align}
where $t$ is the time in which a neutrino event has been measured, or 12:00 UTC of a day when the data is grouped into daily bins.

Two linear regression methods, a daily-binned $\chi^2$ and an unbinned \ac{llh} fit, were independently applied on the same data sample to fit the slope parameter $\alpha$ and verify the robustness of each other. 

Theoretical predictions of $\alpha$ were done by using the software \ac{mceq} \cite{Fedynitch_2017} to numerically calculate the neutrino rates at a given temperature profile measured by \ac{airs}, and then performing pseudo-experiments from the predicted daily rates. For the predictions, the hadronic interaction model SIBYLL2.3c \cite{sibyll2.3c} and the primary \ac{cr} flux model H4a \cite{H4a_2012} were used.

The previous analysis yielded the results and corresponding statistical uncertainties as
\begin{align}
    \label{eq:alpha_theo_south}
    \alpha_\mathrm{theo} & = 0.424\pm0.038(stat.)\pm0.025(sys.) \quad
    \\
    \label{eq:alpha_old_south}
    \alpha_\mathrm{6yr}  & = 0.347\pm0.029 \quad \,,
\end{align}
which are in tension with each other. However, the difference observed does not exceed the $3\sigma$ range and the uncertainty was statistically dominated. The systematic uncertainty on the predicted $\alpha$ originated from the uncertainties in hadronic interactions models and was derived by varying Barr-parameters \cite{nuflux_uncertainty_2006} in the SIBYLL2.3c model. It was shown that these known systematic uncertainties are unable to resolve the tension \cite{6year_paper}.

\section{Updated Datasets}
\begin{figure}
\centering
\includegraphics[width=0.8\linewidth]{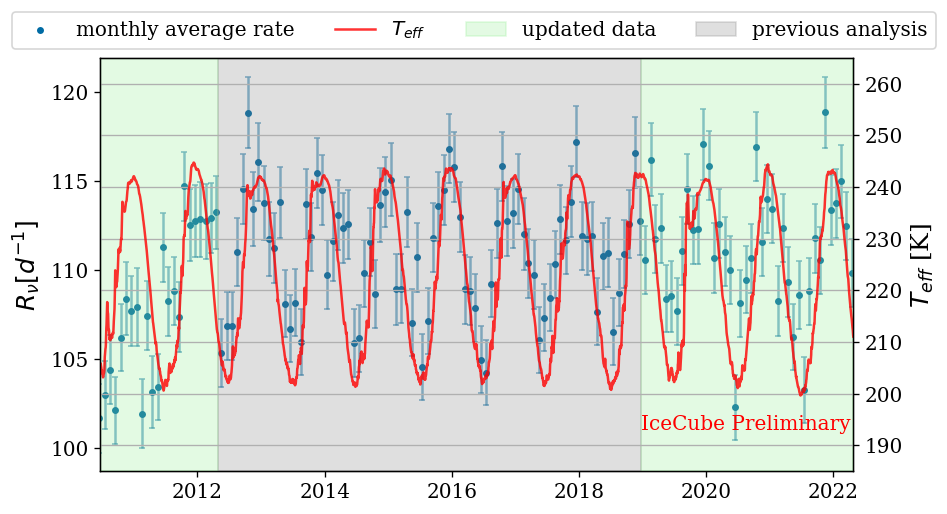}
\caption{Monthly-averaged neutrino rates and effective temperature in the southern observation region over the whole timeframe of the data sample. The gray background denotes the data already used by previous analysis, and the green the updated new data.}\label{fig:south_rates}
\end{figure}
The northern track selection of IceCube is used for this and the previous analysis. It produces a pure high-statistic muon neutrino sample dominated by atmospheric neutrinos \cite{Icecube_munudata_2022}. We expand our timeframe to include all events between June 20, 2010, and April 26, 2022. After the selection, a livetime of 4185.159 days was achieved, corresponding to 96.7\% of the total length of our time range. In total, 982279 events were registered in the sample. Furthermore, we select events with specific reconstructed zenith angles in the IceCube. The zenith region of $90^\circ < \theta < 115^\circ$ was selected for both this and the previous analysis. This zenith range corresponds to the region between the South Pole and the 40°S latitude line, where the temperature fluctuation is most prominent. The effective temperature calculated for this region over the whole analysis time window is shown in Fig.~\ref{fig:south_rates}, along with the monthly averaged neutrino rates measured by IceCube. A correlation between the two variables is visible.

Beside the southern region, the analyzed zenith region has been expanded to also include $155^\circ < \theta < 180^\circ$, corresponding to the region around the geographical North Pole. This region also features a prominent seasonal variation of atmospheric temperature, with a magnitude slightly smaller than the southern region. However, this region has a number of events that is only about 1/10 of the southern region and was therefore not analyzed in the previous work. However, with 12 years of data, we expect enough statistics to resolve seasonal variation in this region. 

The measurement in the Northern Hemisphere also serves as an important cross-check for potential systematic effects. The temperature variation in the Northern Hemisphere is in the opposite phase to that in the Southern, so that unmodeled temperature-independent systematic effects are also expected to have the opposite effect on the measurement. Furthermore, the atmospheric conditions in the North Pole are very different from the South, making this cross-check sensitive to systematic effects caused by unmodeled atmospheric conditions. 
The operation of \ac{airs} continued during the taking of the updated neutrino sample \cite{aqua_2022}, therefore, its temperature measurement is further used in this work.

\section{Results}\label{sec2}

\begin{figure}
\centering
\includegraphics[width=0.8\linewidth]{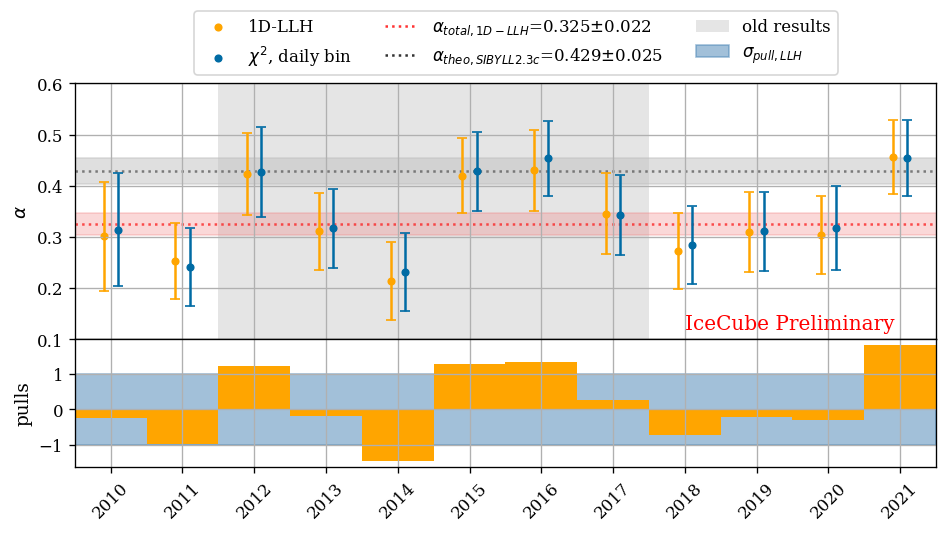}
\caption{Slope parameter $\alpha$ measured with all data is shown as a red dashed line, and the theoretical prediction as a gray dashed line; the corresponding color band indicates the (expected) statistical uncertainties. $\alpha$ fitted with all individual years are shown, along with their pulls in the lower panel. The corresponding color band indicates the standard deviation of the pulls, whose width is close to 1, indicating the difference between each year can be explained with statistical fluctuation.}\label{fig:yearly}
\end{figure}

In our two analysis regions, we obtained from 12 years of data
\begin{align}
    \label{eq:alpha_south}
    \alpha_\mathrm{south} & = 0.325\pm0.022 \quad
    \\
    \label{eq:alpha_north}
    \alpha_\mathrm{north}  & = 0.731\pm0.222 \quad \,,
\end{align}
respectively. The result in the southern region as well as $\alpha$ values measured with data from each individual year are shown in 
Fig.~\ref{fig:yearly}. To quantify the fluctuation through years, we calculate the pull of each individual year, which is defined as
\begin{equation}
    \label{eq:pull}
    \mathrm{pull}_i = \frac{\alpha_i-\left<\alpha_\mathrm{tot.}\right>}{\sigma_i\mathrm{(stat.)}} \,.
\end{equation}
The results are shown in the lower panel of Fig.~\ref{fig:yearly}, along with their standard deviation, which is close to 1. This indicates that the difference between each year agrees with the statistical fluctuation.

\begin{figure}
\centering
\includegraphics[width=0.8\linewidth]{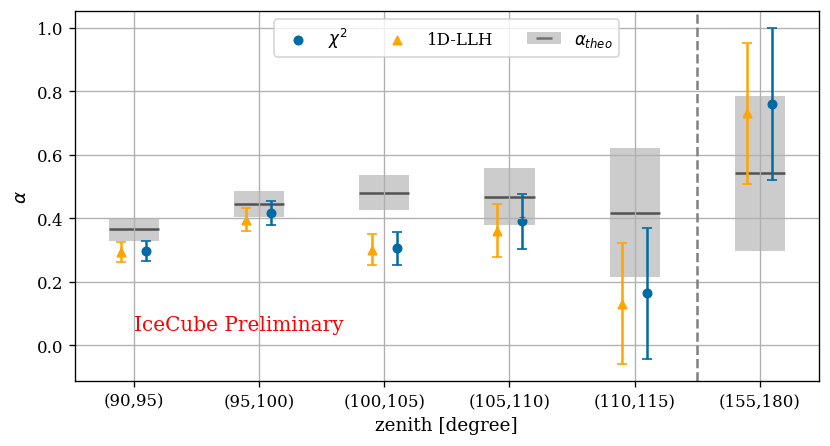}
\caption{Measured and predicted $\alpha$ values in each subdivided zenith bin are shown, as well as the northern observation region. No unexpected zenith-dependent systematic effects are observed.} \label{fig:zenith}
\end{figure}

The result from subdivisions of zenith regions are shown in Fig.~\ref{fig:zenith}, along with the corresponding theoretical prediction. Beside the observed tension between the measured and predicted $\alpha$ value, no unexpected zenith-dependent effects are observed. The measurement of the slope parameter $\alpha$ in the Northern Hemisphere is also shown in Fig.~\ref{fig:zenith} as the rightmost data point.

\begin{figure}
\centering
\includegraphics[width=0.8\linewidth]{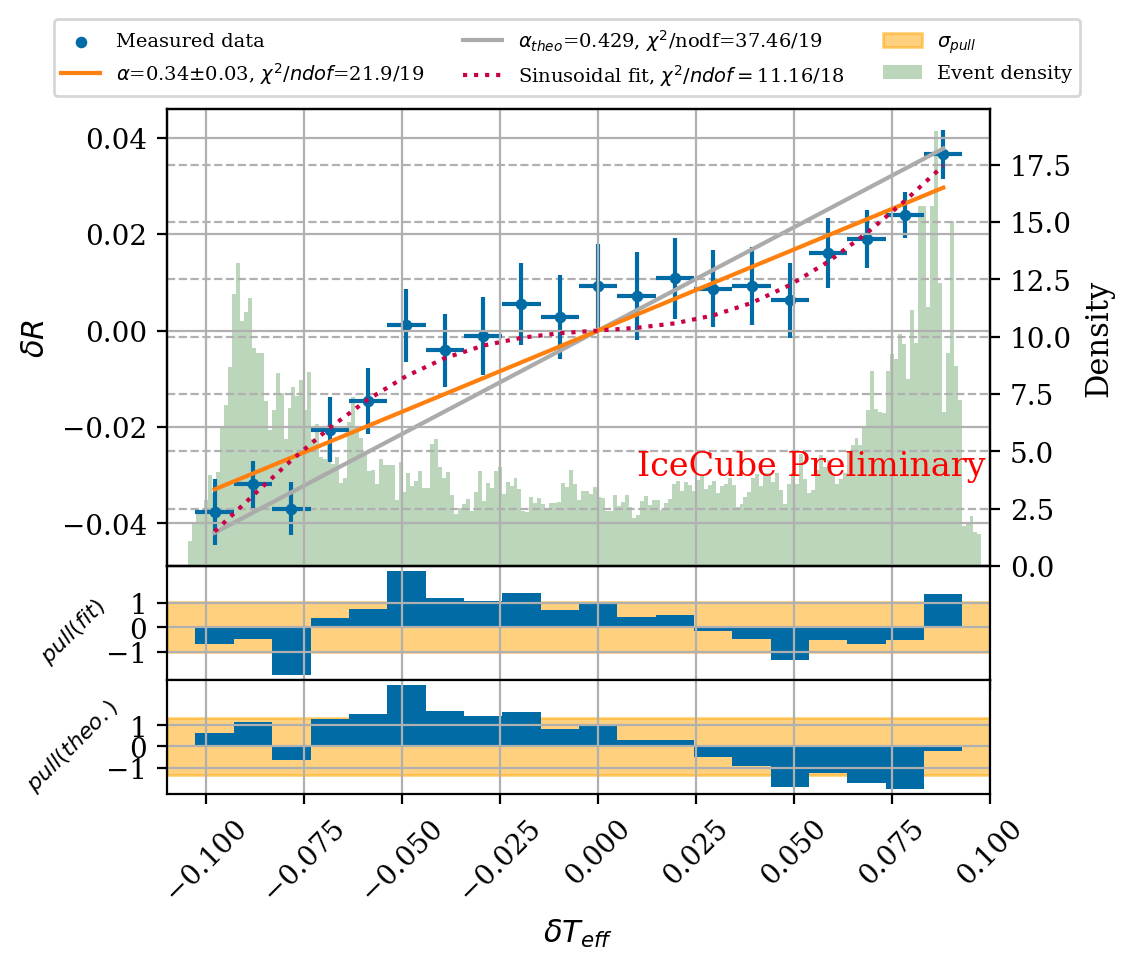}
\caption{The neutrino events are regrouped into temperature bins, and a sinusoidal fit was performed alongside the expected linear fit to demonstrate the systematic effect observed. The green histogram indicates the density of neutrino events observed in the corresponding temperature range. Most of the events are registered during extreme temperatures, since they last longer in regions around the Poles. The two panels on the bottom show the pulls of measured relative rate to the fitted line and the theoretical expectation, respectively. A systematical deviation from the linear expectation is visible.}\label{fig:salsa}
\end{figure}

To investigate the potential systematic effects in the measurement, we regrouped the measured neutrino events according to their corresponding effective temperature. The result is demonstrated in Fig.~\ref{fig:salsa}. To quantify the observed systematic effect, we define a sinusoidal model

\begin{equation}
    \label{eq:salsa_fit}
    \delta R_\nu = \left( \alpha + \beta \sin{\frac{2\pi}{\Delta T_\mathrm{eff,max}}} \right)\delta T_\mathrm{eff}
\end{equation}
with an extra parameter $\beta$, and perform a likelihood-ratio test \cite{Wilks_1938} with the expected linear model as the null hypothesis. As a result, the linear model is disfavored with a p-value of $0.001$. This result indicates the presence of unknown temperature-related systematic effects.

\section{Conclusion}\label{sec3}

IceCube has observed the seasonal variation of atmospheric neutrino flux in a previous measurement in the southern region ($90^\circ < \theta < 115^\circ$) with a resulting slope parameter of $\alpha_\mathrm{6yr} = 0.347\pm0.029$, which was in tension with the theoretical prediction. Possible candidates for the cause of this tension include unexpected uncertainties in the hadronic interaction models, systematic effects in the temperature measurement, and unmodeled detector effects. The previous study included investigations on systematic effects \cite{6year_paper}, while the origin of the tension was not concluded.

In this work, we measured $\alpha$ with 12 years of IceCube data, doubling the statistic in comparison to the previous analysis \cite{6year_paper}. This yielded in this region a measurement of $\alpha=0.325\pm0.022$, which is compatible with the previous result. The previously observed tension between the measurement and the prediction was confirmed with significance $>3\sigma$.

The seasonal variation of atmospheric neutrino flux in the Northern Hemisphere ($155^\circ < \theta < 180^\circ$) was observed with $>3\sigma$ significance for the first time. The slope parameter in the northern region was measured as $\alpha_\mathrm{north} = 0.731\pm0.222$, which is consistent with the theoretical prediction. However, considering the fact that we only have about 1/10 of the statistics compared to the southern region, a systematic effect of similar magnitude to the southern region cannot be excluded.

In this work, we applied an extra statistical test on the temperature-binned neutrino rates and exhibited a non-linear temperature dependency of the tension. Compared to the southern region, the northern region has an opposite temperature phase, a different incline angle for the production of \acp{eas}, and a different zenith angle in the detector for reconstruction. Thus, the successful observation with a result compatible with the prediction disfavors uncertainties on the temperature measurement, mis-modeling of \ac{eas} and reconstruction bias as candidates for the origin of observed tension.

\providecommand{\href}[2]{#2}\begingroup\raggedright\endgroup

\clearpage

\section*{Full Author List: IceCube Collaboration}

\scriptsize
\noindent
R. Abbasi$^{16}$,
M. Ackermann$^{63}$,
J. Adams$^{17}$,
S. K. Agarwalla$^{39,\: {\rm a}}$,
J. A. Aguilar$^{10}$,
M. Ahlers$^{21}$,
J.M. Alameddine$^{22}$,
S. Ali$^{35}$,
N. M. Amin$^{43}$,
K. Andeen$^{41}$
% ...
\\
\\
% ...
$^{10}$ Universit{\'e} Libre de Bruxelles, Science Faculty CP230, B-1050 Brussels, Belgium \\
% ...
$^{16}$ Department of Physics, Loyola University Chicago, Chicago, IL 60660, USA \\
$^{17}$ Dept. of Physics and Astronomy, University of Canterbury, Private Bag 4800, Christchurch, New Zealand \\
% ...
$^{21}$ Niels Bohr Institute, University of Copenhagen, DK-2100 Copenhagen, Denmark \\
$^{22}$ Dept. of Physics, TU Dortmund University, D-44221 Dortmund, Germany \\
% ...
$^{35}$ Dept. of Physics and Astronomy, University of Kansas, Lawrence, KS 66045, USA \\
% ...
$^{39}$ Dept. of Physics and Wisconsin IceCube Particle Astrophysics Center, University of Wisconsin{\textemdash}Madison, Madison, WI 53706, USA \\
% ...
$^{41}$ Department of Physics, Marquette University, Milwaukee, WI 53201, USA \\
% ...
$^{43}$ Bartol Research Institute and Dept. of Physics and Astronomy, University of Delaware, Newark, DE 19716, USA \\
% ...
$^{63}$ Deutsches Elektronen-Synchrotron DESY, Platanenallee 6, D-15738 Zeuthen, Germany \\
$^{\rm a}$ also at Institute of Physics, Sachivalaya Marg, Sainik School Post, Bhubaneswar 751005, India \\
% ...

\subsection*{Acknowledgments}

\noindent
The authors gratefully acknowledge the support from the following agencies and institutions:
USA {\textendash} U.S. National Science Foundation-Office of Polar Programs,
U.S. National Science Foundation-Physics Division,
U.S. National Science Foundation-EPSCoR,
U.S. National Science Foundation-Office of Advanced Cyberinfrastructure,
Wisconsin Alumni Research Foundation,
Center for High Throughput Computing (CHTC) at the University of Wisconsin{\textendash}Madison,
Open Science Grid (OSG),
Partnership to Advance Throughput Computing (PATh),
Advanced Cyberinfrastructure Coordination Ecosystem: Services {\&} Support (ACCESS),
Frontera and Ranch computing project at the Texas Advanced Computing Center,
U.S. Department of Energy-National Energy Research Scientific Computing Center,
Particle astrophysics research computing center at the University of Maryland,
Institute for Cyber-Enabled Research at Michigan State University,
Astroparticle physics computational facility at Marquette University,
NVIDIA Corporation,
and Google Cloud Platform;
Belgium {\textendash} Funds for Scientific Research (FRS-FNRS and FWO),
FWO Odysseus and Big Science programmes,
and Belgian Federal Science Policy Office (Belspo);
Germany {\textendash} Bundesministerium f{\"u}r Forschung, Technologie und Raumfahrt (BMFTR),
Deutsche Forschungsgemeinschaft (DFG),
Helmholtz Alliance for Astroparticle Physics (HAP),
Initiative and Networking Fund of the Helmholtz Association,
Deutsches Elektronen Synchrotron (DESY),
and High Performance Computing cluster of the RWTH Aachen;
Sweden {\textendash} Swedish Research Council,
Swedish Polar Research Secretariat,
Swedish National Infrastructure for Computing (SNIC),
and Knut and Alice Wallenberg Foundation;
European Union {\textendash} EGI Advanced Computing for research;
Australia {\textendash} Australian Research Council;
Canada {\textendash} Natural Sciences and Engineering Research Council of Canada,
Calcul Qu{\'e}bec, Compute Ontario, Canada Foundation for Innovation, WestGrid, and Digital Research Alliance of Canada;
Denmark {\textendash} Villum Fonden, Carlsberg Foundation, and European Commission;
New Zealand {\textendash} Marsden Fund;
Japan {\textendash} Japan Society for Promotion of Science (JSPS)
and Institute for Global Prominent Research (IGPR) of Chiba University;
Korea {\textendash} National Research Foundation of Korea (NRF);
Switzerland {\textendash} Swiss National Science Foundation (SNSF).

\end{document}